\documentstyle[12pt,epsfig]{article}
\newcommand{\beq}{\begin{equation}}
\newcommand{\eeq}{\end{equation}}
\newcommand{\bea}{\begin{eqnarray}}
\newcommand{\eea}{\end{eqnarray}}

\newcommand{\gsim}{\lower.7ex\hbox{$
\;\stackrel{\textstyle>}{\sim}\;$}}
\newcommand{\lsim}{\lower.7ex\hbox{$
\;\stackrel{\textstyle<}{\sim}\;$}}

\newcommand{\eod}{\end{document}}

\def\op{{\bf P}}

\def\ot{{\bf T}}
\def\cp{{\bf CP}}
\def\cpt{{\bf CPT}}

\begin{document}
\thispagestyle{empty}
\vspace*{-22mm}

\begin{flushright}
UND-HEP-09-BIG\hspace*{.08em}01\\

\end{flushright}
\vspace*{1.3mm}

\begin{center}
{\Large {\bf
Could Charm's ``Third Time'' Be the Real Charm? -- \\
\vspace{4mm}
A Manifesto
}}
\vspace*{19mm}

{\bf I.I.~Bigi} \\
\vspace{7mm}

 {\sl Department of Physics, University of Notre Dame du Lac}
\vspace*{-.8mm}\\
{\sl Notre Dame, IN 46556, USA}\\
{\sl email: ibigi@nd.edu} \\

\vspace*{10mm}

{\bf Abstract}\vspace*{-1.5mm}\\
\end{center}

\noindent
The recent observation of $D^0 - \bar D^0$ oscillations has left us in a quandary concerning the theoretical 
interpretation: are they driven by SM forces alone or do they involve new dynamics? A comprehensive 
analysis of \cp~symmetry in $D$ decays can probably resolve the issue. Charm studies 
might thus haul in their biggest prize yet:  clear evidence for the intervention of New Physics.  
While the absolute size of \cp~asymmetries will presumably be modest at best, SM contributions should be much smaller still. Thus the ratio of `signal' to `noise'  -- i.e. NP over SM contributions -- 
might well be larger for $D$ than $B$ transitions.  
A typical list of promising channels is provided, most of which should be observable in a hadronic environment. Valuable lessons can be obtained by analyzing three- and four-body final states.  

\tableofcontents

\vspace{1.5cm}

\section{Prologue on Charm and its Uniqueness}

Charm's ``first time" was instrumental in the paradigm shift concluded by the ``October" or 
``$J/\psi$ Revolution" of 1974 
\footnote{It is usually called the {\em November Revolution}, since the discovery was announced on Nov. 11, 1974. My terminology stresses the revolutionary aspect by its obvious reference to the 1917 Russian Revolution which according to the calendar in force in Russia took place in October; according to our modern calendar the storming of the Winter palace occurred Nov. 7, 1917. There is another and deeper analogy: While this storming was a dramatic event, the real change was caused by events that had happened some months earlier. }
that validated quarks as truly physical degrees of freedom. The 
``second time" it reminded us of our lack of knowledge concerning QCD's nonperturbative dynamics, when the $D_{sJ}$ resonances were found. The observation of $D^0 - \bar D^0$ oscillations in 2007 
\cite{BABAROSC,BELLEOSC} might be the first step towards the ``third time", when forces beyond the Standard Model (SM) are revealed in 
charm transitions, which would be the greatest discovery with charm. 

New Physics (NP) will in general induce flavour changing neutral currents (FCNC). The SM had to be crafted judiciously to have them suppressed sufficiently for strangeness; the weight of FCNC is then even more reduced for the up-type quarks $u$, $c$ and $t$. Yet this does not 
hold in general: FCNC in NP scenarios could be more relevant for up-type quarks. Among those   
it is only the charm quark that allows the full range of probes for FCNC in general and for 
\cp~violation in particular. For top quarks do not hadronize \cite{RAPALLO} thus eliminating the occurrence of 
$T^0 - \bar T^0$ oscillations. Neutral pions etc. cannot oscillate, since they are their own antiparticles; furthermore \cpt~constraints are such that they rule out most \cp~asymmetries.  
Yet now we know that neutral charm mesons oscillate and roughly at which rate, namely characterized 
by $x_D \equiv \Delta M_D/\overline \Gamma_D$ $\sim$ 
$y_D \equiv \Delta \Gamma_D/2\overline \Gamma_D$ $\sim 0.005 - 0.01$ \cite{HFAG}. 

The theoretical interpretation is unclear: while the SM can 
`naturally' generate $x_D \sim y_D \sim {\cal O}(10^{-3})$, one cannot rule out values as `large' 
as 0.01 \cite{CICERONE,FALK}. Despite the similar numerical estimates for $x_D$ and $y_D$ the underlying dynamics is of a very different nature: while $\Delta M_D$ is generated with {\em off}-shell intermediate states, $\Delta \Gamma_D$ is obtained from {\em on}-shell ones; 
$\Delta M_D$ can thus naturally be sensitive to  New Physics, which is unlikely for $\Delta \Gamma_D$, 
though not impossible \cite{PETROV}. 

Finding $x_D \gg y_D \sim {\cal O}(10^{-3})$ would have represented strong prima facie evidence for 
the presence of NP. Such a scenario appears to have been ruled out. The present situation can instead be interpreted in two ways: (i) It is beyond our computational abilities to evaluate 
$\Delta M_D$ and $\Delta \Gamma_D$ accurately. (ii) It represents one example of nature being mischievous with us: $\Delta \Gamma_D$ is anomalously enhanced due to a violation of 
{\em local} quark-hadron duality caused by the proximity of hadronic thresholds 
\cite{CICERONE,DUA}; 
$\Delta M_D$ on the other hand is enhanced by NP over the value expected in the SM. 

Interpretation (i) presumably reflects the majority opinion despite (or because ?) it has to be seen as a blow to the professional pride of us theorists. Interpretation (ii) on the other hand would restore it. 
The latter will not be validated by theoretical arguments alone. Yet there is a straightforward course of action to clarify and hopefully decide the issue, namely to conduct a comprehensive and dedicated study of \cp~invariance in charm decays. One cannot count on NP creating larger effects in $D$ than in $B$ decays and their \cp~asymmetries, but their manifestations might be clearer in the former than the latter; for the SM induces still much smaller effects: 
\beq 
\left[ \frac{\rm experim. \; NP\; signal}{\rm SM \; \cp~``background"}\right]_{{\bf D}\; {\rm decays}} > 
\left[ \frac{\rm experim. \; NP\; signal}{\rm SM \; \cp~``background"}\right]_{{\bf B}\; {\rm decays}}
\eeq
In summary: while the observed signal for $D^0 - \bar D^0$ oscillations represents a tactical draw in 
our struggle to reach beyond the SM, there is new promise for a strategic victory on the battleground 
of \cp~studies. It is the relative `dullness' of the SM vis-a-vis \cp~asymmetries in charm transitions that 
make such searches promising. The observation of oscillations has widened the stage where NP can reveal itself.  

There is a qualitative analogy with the case of $B_s - \bar B_s$ oscillations: the observed rate as expressed through $\Delta M_{B_s}$ is fully consistent with SM predictions \cite{HFAG} -- within sizable theoretical 
uncertainties. The next challenge is to search for a time dependent \cp~asymmetry in 
$B_s \to \psi \phi$. For KM dynamics predicts \cite{BS80} a Cabibbo suppressed 
asymmetry $\sim 3$ \%. NP whose non-leading contribution to $\Delta  M_{B_s}$  
might be hiding behind the theoretical uncertainty in the SM prediction could enhance the \cp~asymmetry even by an order of magnitude thus becoming the 
{\em leading} effect {\em there}. 

In Sect.\ref{OVER} I will give an overview before addressing some of the technical points in 
more detail in Sect.\ref{TECH} and presenting an outlook in Sect.\ref{OUT}. 

\section{Overview}
\label{OVER}

\subsection{Fundamentals of \cp~Searches and NP Scenarios}
It is doubtful that the SM predictions for $x_D$ and $y_D$ can be refined  significantly in the near future. 
Nevertheless determining $x_D$ and $y_D$ with good accuracy is motivated by pragmatic 
rather than quixotically noble reasons: a measurement of a presumably small time dependent \cp~asymmetry would be validated by reproducing values for $x_D$ and $y_D$ consistent with independent data. Furthermore those accurate values are needed to identify the source(s) of an asymmetry, whether it is 
due to $\left| \frac{q}{p}\right| \neq 1$ or arg$\frac{q}{p}\bar \rho_f \neq 0$, see Sect.\ref{LIST}. 

Searching for manifestations of NP through \cp~studies is not a `wild goose' chase, since baryogenesis 
requires the intervention of NP {\em with} \cp~violation; it is not an unreasonable hope that such 
\cp~odd NP would leave its mark also somewhere else -- like in charm transitions. One should note that a \cp~asymmetry is {\em linear} rather than quadratic in a NP amplitude, which enhances the 
sensitivity to small amplitudes. 

One drawback in searching for NP in charm decays is the fact that those occur already on the Cabibbo allowed level unlike for kaons and $B$ mesons, whose modes are Cabibbo and KM  suppressed, respectively. Since it is unlikely that NP in Cabibbo favoured channels would have escaped discovery, one has to search for it in significantly suppressed modes. Acquiring large data sets is one ingredient for overcoming this challenge. 

Otherwise most 
{\em experimental}  features favour the observability of \cp~asymmetries in charm decays: 
(i) The effective branching ratios into pions, kaons and leptons for many relevant modes are relatively sizable. 
(ii) Final state interactions (FSI) needed to make direct \cp~violation observable in two-body final states are generally 
large \footnote{Our lack of theoretical control over final state interactions becomes a problem only when 
{\em interpreting} observations in terms of the {\em microscopic} parameters of the 
underlying dynamics.}. 
(iii) Flavour tagging can conveniently be done by the soft pions in $D^{*\pm} \to D/\bar D \pi^{\pm}$. 
(iv) Many nonleptonic  final states contain more than two pseudoscalar or 
one pseudoscalar and one vector meson. Final state {\em distributions} are then {\em non}trivial and can exhibit 
\cp~asymmetries, which can be studied through Dalitz plots and \ot~odd moments. 
(v) The observation of 
$D^0 - \bar D^0$ oscillations greatly widens the stage for \cp~asymmetries, since it provides a second coherent, yet different amplitude, the weight of which changes with the time of decay. 

On the {\em phenomenological} side there are promising features as well. Since 
$D^0 \to \bar D^0$ transitions are so suppressed within the SM, they provide a promising portal for 
NP to enter in a perceptible way. $\Delta C=1$ nonleptonic modes occur on three Cabibbo levels -- allowed (CA), singly (SCS) and doubly Cabibbo suppressed (DCS) -- whose typical rates differ by tg$^2\theta _C$ and tg$^4\theta _C$, i.e. by one or two to three orders of magnitude, respectively. It is not unreasonable that NP could affect the decay amplitude for DCS and possibly even for SCS modes. Furthermore the SM provides not merely a classification -- it makes highly nontrivial, even if not always accurate predictions concerning \cp~asymmetries: no {\em direct} \cp~violation can occur in CA and DCS channels (except for final states containing $K_S$ [or $K_L$] mesons); it can for SCS modes, but only on a fairly tiny level, since the required weak phase is highly diluted to the tune of 
${\cal O}( {\rm tg}^4\theta _C) \leq 0.001$. In the SM model  
the oscillation amplitude is expected to carry a tiny weak phase 
$\sim {\cal O}({\rm tg}^4\theta _C) \leq 0.001$ 
as a benchmark figure \cite{CICERONE}. Even a NP contribution that is non-leading in $\Delta M_D$ can thus easily provide not only the leading source for \cp~violation, but even a sizable one.  

The last comment can be supported by an explicit model of `nearby' NP, namely the so-called Littlest Higgs model with T parity (LHT) \cite{BLANKE}. Little Higgs models in general provide scenarios  
where NP quanta can be produced at the LHC without creating conflicts with electroweak constraints; 
i.e. they are {\em not} motivated by considerations concerning flavour dynamics. It turns out that LHT dynamics can naturally generate the observed value of $\Delta M_D$ or some fraction of it. Yet then it would most likely induce also a weak phase in $D^0 - \bar D^0$ oscillations that conceivably could be as large as 10 - 15 degrees \cite{BBBR}.   

\subsection{Cast of Candidate Channels}
\label{CAST}
Here I give a list of relevant or even promising channels, most of which -- but not all -- should be observable in hadronic collisions: 
\bea
D^0(t) &\to& K_S \phi , \; K_SK^+K^-, \; K_S \rho , \; K_S\pi^+\pi^- , \;  
K_S \eta ^{(\prime)}, \; K_S \omega \\
D^0(t) &\to & l^- \bar \nu K^+ \\
D^0(t) &\to& K^+K^-, \; \pi^+\pi^-, \; K^+\pi^- \\
D^0(t) &\to & 3\pi , \; K\bar K\pi \\ 
D^0(t) &\to & K^+K^-\pi^+\pi^-, \; K^+K^-\mu^+\mu^-, \; K^+\pi^-\pi^+\pi^- , \; K^+K^-K^+\pi^-\\
D^{\pm} &\to& K_S\pi^{\pm}, \; K_S K^{\pm} \\
D^{\pm} &\to & 3\pi , \; K\bar K\pi, \; K^{\pm}K^+K^-, \;  K^{\pm}\pi^+\pi^-\\ 
D^0 &\to& \mu^+\mu^- \, [,\, \gamma \gamma]
\eea
It should be noted that this is at best a representative list, not an exhaustive one. The dedicated 
reader is invited to come up with her/his personal favourite; in particular she/he can identify 
the corresponding $D_s$ modes. 

\subsection{List of Relevant Observables}
\label{LIST} 

There are two a priori distinct portals for \cp~violation: it can enter via $\Delta C=1$ 
or $\Delta C=2$ dynamics; in the former [latter] it is called -- with less than Shakespearean 
flourish -- direct [indirect] \cp~violation. 
\begin{itemize}
\item 
{\em Direct} \cp~violation can reveal itself  
through 
a difference in the moduli of the $\Delta C=1$ decay amplitudes describing \cp~conjugate transitions: 
\beq 
|T(D \to f)| \neq |T(\bar D \to \bar f)| \; . 
\eeq
It requires the presence of two coherent amplitudes differing 
in both their weak as well as strong phases. 
\item 
The effects of $D^0 - \bar D^0$ oscillations on \cp~asymmetries can be expressed through 
\beq 
\frac{q_D}{p_D} = \sqrt{\frac{(M_{12}^D)^* - \frac{i}{2}(\Gamma_{12}^D)^*}
{M_{12}^D - \frac{i}{2}\Gamma_{12}^D}}
\eeq
Since $M_{12}^D$ and $\Gamma_{12}^D$ depend on the phase convention chosen for $\bar D^0$, neither of them nor $q_D/p_D$ can be observables by themselves. Yet two types of \cp~observables can be inferred from the latter:
\begin{itemize}
\item 
Using $\Gamma_{12}^D \simeq \Gamma_{12}^D(SM)$ and ignoring the tiny CKM phase in 
$\Gamma_{12}^D(SM)$ while allowing for NP to contribute significantly to 
$M_{12}^D$ and its phase -- $M_{12}^D = |M_{12}^D|e^{i\xi_{\Delta C=2}}$ -- leads to 
$$ 
\left| \frac{q_D}{p_D} \right| ^4 = 1 +  \Delta_4   \neq 1 
$$
\beq
\Delta_4 = \frac{2\Gamma_{12}^D|M_{12}^D| {\rm sin}\xi_{\Delta C=2}}
{|M_{12}^D|^2 +\frac{1}{4} |\Gamma_{12}^D|^2 -\Gamma_{12}^D|M_{12}^D| {\rm sin}\xi_{\Delta C=2}}
\sim \frac{4x_Dy_D{\rm sin}\xi_{\Delta C=2}}{x_D^2 + y_D^2 -2x_Dy_D{\rm sin}\xi_{\Delta C=2}}
\label{DELTA4}
\eeq
$|q_D/p_D| \neq 1$ unequivocally describes {\em indirect}  \cp~violation in $D^0 - \bar D^0$ 
oscillations,  
and it affects semileptonic and the 
appropriate nonleptonic channels. 
\item 
When a nonleptonic final state is common to $D^0$ and $\bar D^0$ decays -- in the simplest such 
cases it will be a \cp~eigenstate --  it can exhibit a time-dependent asymmetry 
described by 
\beq 
{\rm sin}(\Delta M_D t){\rm Im}\frac{q_D}{p_D} \frac{T(\bar D^0 \to f)}{T(D^0 \to f)} 
\simeq x_D \frac{t}{\tau_D} {\rm Im}\frac{q_D}{p_D} \frac{T(\bar D^0 \to f)}{T(D^0 \to f)} \; , 
\eeq
which reflects the combined effect of oscillation and direct decay; it is driven again by 
$\xi_{\Delta C=2} \neq 0$ and affected by $M_{12}^D$, $\Gamma_{12}^D$ and phases in the 
decay amplitudes for $D^0/\bar D^0 \to f$. 
\item 
Oscillations can provide access to {\em direct} \cp~violation that might otherwise remain unobservable. 
Consider two different final states $f_1$ and $f_2$ that are \cp~eigenstates and thus 
common to $D^0$ and $\bar D^0$ decays; for example $f_1 = \phi K_S$ and $f_2=K^+K^-$. 
In the absence of direct \cp~violation one has 
\beq 
{\rm Im} \frac{q}{p} \frac{T(\bar D^0 \to f_1)}{T(D^0 \to f_1)}\; = \; \eta_{f_1f_2} 
{\rm Im} \frac{q}{p} \frac{T(\bar D^0 \to f_2)}{T(D^0 \to f_2)} 
\eeq
with $\eta_{f_1f_2}$ denoting the relative \cp~parity of $f_1$ vs. $f_2$; it is -1 in the example given above. Any difference from this relation shows \cp~violation in at least one of the $\Delta C=1$ amplitudes. It should be noted that in the absence of oscillations this source of \cp~violation would be 
unobservable if the $D\to f_1$ and/or $D\to f_2$ amplitudes did not each contain two weak and 
two strong phases. 

\end{itemize}

\end{itemize}
I will list here channels that appear to be most promising for revealing such effects on various Cabibbo levels; the observables are partial rates, Dalitz plot parameters and \ot~odd moments. Again no claim is made for this being a complete list -- the reader is invited to come up with her/his favourite modes. 

\pagebreak

\subsubsection{Two-body Final States}

{\bf Cabibbo allowed modes}

Searching for asymmetries in CA final states of neutral $D$ mesons represents a clean search 
for indirect \cp~violation, since NP affecting CA $\Delta C =1$ amplitudes should have been noted by now. The theoretically simplest channels would be 
\beq 
D^0 \to K_S\pi^0, \, K_S\eta, \, K_S \eta^{\prime}
\eeq 
-- alas experimentally they 
are anything but simple. In a hadronic environment they seem to be close to impossible. The next best 
mode is
\beq 
D^0 \to K_S\phi \to K_S[K^+K^-]_{\phi} \; , 
\eeq
which is given by a single 
isospin amplitude. The {\em strong} phase thus drops out from the ratio 
$\bar \rho_{K_S\phi} \equiv \frac{T(\bar D^0 \to K_S\phi)}{T(D^0 \to K_S\phi)}$, while their 
tiny SM weak 
phase can be ignored at first. Using the notation 
$\frac{q_D}{p_D}\bar \rho (K_S\phi) = (1+\Delta _D) e^{i\xi^{NP}_{\Delta C=2}}$ and ignoring contributions 
higher than linear in $x_D$ and $y_D$ one then finds  
\bea 
\frac{\Gamma(D^0(t) \to K_S\phi) - \Gamma(D^0(t) \to K_S\phi)}
{\Gamma (D^0(t) \to K_S\phi) + \Gamma(D^0(t) \to K_S\phi)} &\simeq &
- (y_D\Delta_D {\rm cos}\xi^{NP}_{\Delta C=2} +
 x_D {\rm sin}\xi^{NP}_{\Delta C=2} )\frac{t}{\overline \tau _D} \\
&\simeq & - (y_D\Delta_D + x_D \xi^{NP}_{\Delta C=2} )\frac{t}{\overline \tau _D} \; , 
\label{KSPHI}
\eea
where the last simplified expression holds for $\xi^{NP}_{\Delta C=2} \ll 1$. 
We have here a {\em qualitative}  analogy to $B_d \to \psi K_S$. The effect will be much smaller of course with 
much slower oscillations, and a priori one cannot ignore the impact of $y_D \neq 0$ and 
$\left| \frac{q}{p} \right| \neq 1$. The expression also exemplifies a statement made above: 
having an independent and accurate measurement of $y_D$ and $x_D$ provides great help for  
properly interpreting an asymmetry in terms of $\Delta _D$ vs. $\xi^{NP}_{\Delta C=2}$. 

Extracting $D^0 \to K_S \phi$ from $D^0 \to K_SK^+K^-$ is not a trivial undertaking. Among other challenges one has to distinguish it 
from $D^0 \to K_Sf^0$. For the \cp~parities of 
$K_Sf^0$ and $K_S\phi$ are opposite. Therefore these final states would have to exhibit \cp~asymmetries of equal size, yet opposite sign. Ultimately one has to and can perform a 
\cp~analysis of the full Dalitz plot for $K_SK^+K^-$; describing it is beyond the scope of this note. 
It should be noted though that this Dalitz analysis is for validation purposes only rather than to search for  
sources of direct \cp~violation. For I assume it to be absent on the CA level; below I will describe one exception to this assumption. 

Another suitable and actually multi-layered channel is 
\beq 
D^0 \to K_S\pi^+\pi^-
\eeq 
As in the previous example one starts 
with the resonant final states $D^0 \to K_S\rho^0$ and then proceeds to a Dalitz plot analysis. There is 
more complexity (pun intended) in this final state: In addition to resonant $\pi^+\pi^-$ states one has 
CA and DCS $K^*\pi$ configurations, namely $D^0 \to K^{*-}\pi^+$ and $K^{*+}\pi^-$, respectively; 
furthermore one has to deal with more than one isospin 
amplitude in the Dalitz plot analysis.  Yet this richer structure also offers more dynamical information that can be inferred from the data and provide more validation of a signal. 

The mode 
\beq
D^{\pm} \to K_S \pi^{\pm} 
\eeq
at first sight appears to be a CA mode. However that final state can be reached also through a DCS amplitude -- 
$D^{\pm} \to \overline K^0 \pi^{\pm} / K^0 \pi^{\pm} \to K_S \pi^{\pm}$ -- and its interference with the 
CA amplitude provides a non-negligible contribution. With{\em out} NP there are already two sources for a direct \cp~asymmetry \cite{YAMA}: (i) The interference between 
$D^{\pm} \to \overline K^0\pi^{\pm} \to K_S\pi^{\pm}$ and 
$D^{\pm} \to K^0\pi^{\pm} \to K_S\pi^{\pm}$ generates an asymmetry 
$\sim {\cal O}({\rm tg}^6\theta_C ) \sim 10^{-4}$. (ii) The \cp~impurity in the $K_S$ wave function 
induces a larger asymmetry: 
\beq 
\frac{\Gamma (D^+\to K_S\pi^+) - \Gamma (D^-\to K_S\pi^-)}
{\Gamma (D^+\to K_S\pi^+) + \Gamma (D^-\to K_S\pi^-)} \simeq 
\frac{|q_K|^2 - |p_K|^2}{|q_K|^2 + |p_K|^2} \simeq - (3.32 \pm 0.06) \cdot 10^{-3}
\eeq
Any deviation from this accurate prediction would be due to an intervention by NP, presumably 
entering through the DCS amplitude. 

Semileptonic decays of neutral $D$ mesons do not lead to two-body final states, yet their discussion fits 
into the present context, since their `wrong-sign' lepton rates exhibit an asymmetry 
\beq 
a_{SL}(D^0) \equiv \frac{\Gamma (D^0\to l^-X) - \Gamma (\bar D^0\to l^+X)}
{\Gamma (D^0\to l^-X) + \Gamma (\bar D^0\to l^+X)} \simeq 
\frac{|q_D|^4 - |p_D|^4}{|q_D|^4 + |p_D|^4} \simeq  \frac{\Delta_4}{2} 
\label{ASL}
\eeq
independent of the time of decay, see Eq.(\ref{DELTA4}). The data point to the factor $2x_Dy_D/(x_D^2 + y_D^2)$ being about one. 
It should be noted that at least some NP scenarios can induce a sizable weak phase in 
${\cal L}(\Delta C=2)$; e.g., Littlest Higgs Models with T parity could produce 
$\phi_{weak} \sim (10 - 15) ^{\circ}$ \cite{BBBR}. We know already that the `wrong-sign' rate is very small, namely $\simeq (x_D^2 + y_D^2)/2 \leq (1.3 \pm 2.7) \cdot 10^{-4}$. Yet the {\em asymmetry}  
$a_{SL}(D^0)$ might even be large. This is a situation quite unlike with what happens for oscillating 
neutral mesons built from down-type quarks, namely $K_L$, $B_d$ and $B_s$. There one has plenty 
of wrong-sign leptons, yet $a_{SL}$ is very small at best, but for somewhat different reasons: for 
the neutral $B$ mesons it is suppressed mainly due to $\Delta \Gamma_B \ll \Delta M_B$ and for 
$K_L$ due to the effective decoupling of the third family from the first two. 

While semileptonic $D^0$ decays, which presumably are beyond the reach of LHCb, provide the most direct way to probe $|q_D| \neq |p_D|$, many nonleptonic $D^0$ modes are sensitive to it as well.

\vspace{3mm}

{\bf Once Cabibbo suppressed modes} 

\noindent 
Here the situation becomes more complex, since already CKM dynamics can induce 
\cp~asymmetries -- albeit highly diluted ones to the tune of $10^{-3}$ or less -- since two different 
isospin amplitudes contribute. That opens the door wider for NP to enhance an asymmetry over its tiny 
SM expectation, even when it yields no more than a non-leading contribution to the rate. 

Prime examples for promising channels are 
\beq 
D^0 \to K^+K^-, \; \pi^+\pi^-
\eeq
where now direct as well as indirect \cp~violation can arise. In the absence of the former the time dependent asymmetry has to be the same for both channels, since it is driven by the oscillations 
common to both modes. 

Belle and BaBar have searched for a time-{\em integrated} \cp~asymmetry in those two modes, but found none \cite{PDG08}: 
\beq
A_{\cp}(K^+K^-) = (0.1 \pm 0.5) \cdot 10^{-2} \; \; , \; \; 
A_{\cp}(\pi^+\pi^-) = (0.0 \pm 0.5) \cdot 10^{-2} 
\eeq
Yet for a proper perspective one should note that a time-dependent asymmetry involving 
oscillations is bounded by $x_D$, $y_D$. For $x_D$, $y_D \leq 0.01$, as indicated by the data, 
such an asymmetry could hardly exceed the 1\% range. Yet any improvement in the experimental sensitivity for  
$D^0(t) \to K^+K^-, \, \pi^+\pi^-$ constrains NP scenarios -- or could reveal them 
\cite{GKN}.

Since $D^{\pm} \to \pi^{\pm}\pi^0$ leads to an isospin-two final state, one does not expect a 
\cp~asymmetry there. The situation is much more promising for 
\beq 
D^{\pm} \to K^{\pm}K_S 
\eeq
with $D^{\pm} \to \pi^{\pm}\eta^{(\prime)}$ providing the compensating asymmetry to satisfy \cpt~constraints. 

\vspace{3mm}

{\bf Doubly Cabibbo suppressed modes} 

\noindent 
Some DCS contributions have already been considered, when they enter through the `backdoor' 
of final states containing a $K_S$ (or $K_L$). A promising pure DCS channel 
is \cite{BIBERK,NIRETAL}: 
\beq 
D^0 \to K^+\pi^-
\eeq
\begin{itemize}
\item 
For it allows to track both direct and indirect \cp~violation and separate it through analyzing how the asymmetries evolve with the time of decay. 
\item 
The SM amplitude being DCS is significantly reduced by tg$^2\theta_C \sim 1/20$. 

\end{itemize}
It should be noted that sources of indirect and direct \cp~violation could be quite unrelated to each other. 

\subsubsection{Dalitz Plot Studies}

Final states with three pseudoscalar mesons can be treated in a `Catholic' style: there is a single path to `heaven' provided by the Dalitz plot. The challenge we face here can be summarized as follows: we look for probably smallish asymmetries in subdomains of the Dalitz plot, which is shaped by 
nonperturbative dynamics. Large statistics are necessary, yet not sufficient. As far as pattern recognition is concerned, we can learn a lot from astronomers. They regularly face the problem of searching for something they do not quite know what it is at a priori unknown locations and having to deal with background sources that are all too often not really understood. While this sounds like a hopeless proposition, astronomers have actually been quite successful in overcoming these odds. Inspired by astronomers the Rio group \cite{RIOG} has made an intriguing suggestion: rather than search for the customary asymmetry 
$(N- \bar N)/(N + \bar N)$ in particle vs. anti-particle populations $N$ and $\bar N$, respectively, analyze 
\beq 
             \sigma \equiv \frac{N- \bar N}{\sqrt{N + \bar N}} 
\eeq
It corresponds to standard procedure in astronomy -- suggested in 1983 for gamma ray astronomy and now adopted also by the Auger collaboration --  when comparing on- vs. off-source intensity. For a Poissonian distribution the standard deviation can be written as 
$\sigma = \frac{N_{\rm on}- \alpha N_{\rm off}}{\sqrt{N_{\rm on} + \alpha N_{\rm off}}}$. This is just one possible example for how we can learn from our astronomer colleagues -- I am sure there will be more.  

While a proper Dalitz analysis requires a considerable `overhead' in setting it up, it offers -- like 
\ot~{\em odd} correlations addressed next -- valuable `pay-offs': 
\begin{itemize}
\item 
Local asymmetries are bound to be larger than integrated ones. 
\item 
There are correlations in a Dalitz plot that a proper analysis has to exhibit. Such correlations 
provide us with powerful validation tools in particular for smallish effects. 
\item 
Last, but certainly not least, \cp~asymmetries in a Dalitz plot can provide us with information about the 
underlying operators -- their Lorentz and chirality structure etc. -- that are not revealed in two-body modes. 

\end{itemize}

\subsubsection{\ot~{\em Odd} Correlations}

Going beyond three-body final states one has to deal with a `Calvinist' situation. {\em A priori} there are several paths to heaven, and heaven's blessing is revealed {\em a posteriori} by the success of one's 
efforts; i.e. which distribution will provide the clearest \cp~asymmetry depends on the specifics of the underlying dynamics. 

Since under \ot~both momenta $\vec p$ and spin vectors $\vec s$ change sign, the most elementary \ot~odd 
moments are given by expectation values of triple correlations like 
\beq 
\langle \vec p_1\cdot (\vec p_2 \times \vec p_3 \rangle \; \; \; {\rm and/or} \; \; \; 
\langle \vec s\cdot (\vec p_1 \times \vec p_2 \rangle
\eeq
Unless one has access to spin vectors one obviously needs at least a four-body final state for a 
\ot~odd moment. 

There is a subtle, yet important distinction between \ot~odd and, say, \op~odd moments. Observing a 
\op~odd moment unequivocally establishes \op~violation unlike for the case of a \ot~odd moment. 
For the latter can be generated also with \ot~invariant dynamics if one goes beyond lowest order; 
i.e. FSI can {\em fake} a \ot~violation. This complication is due to time reversal being described by 
{\em anti}linear transformations. While FSI are a necessary evil for \cp~asymmetries to emerge in partial rates, they can be a nuisance for \ot~odd effects: while they are not needed, they can fake an effect.  

There are basically two ways to deal with this interpretative challenge: (i) One can attempt to estimate the order of magnitude of such FSI effects. (ii) One can compare \ot~odd moments in  \cp~conjugate  
decays of particles and antiparticles: If they are not equal in magnitude, yet opposite in sign, 
\ot~invariance is broken, since \cp~transformations are linear. 

The simplest cases are provided by 
\bea
D^0 \to K^+K^- \pi^+\pi^-      \; \; \;& vs. &\; \; \; \bar D^0 \to K^+K^- \pi^+\pi^-   \\
D^0 \to K^+K^- \mu^+\mu^-      \; \; \;& vs. &\; \; \; \bar D^0 \to K^+K^- \mu^+\mu^-   \\
D^+ \to K^+K_S \pi^+\pi^-      \; \; \;& vs. &\; \; \;  D^- \to K^-K_S \pi^+\pi^-  \; , 
\eea
since all particles in the final state are distinct. It should be emphasized again that these are merely 
the most straightforward channels. One can also use, say, $D^0 \to \pi^+\pi^- \pi^+\pi^-$ 
and use selection criteria like the energies of the pions.  It also makes sense to analyze \ot~odd 
moments for neutral $D$ decays as a function of the (proper) time of decay, since oscillations 
can affect the relative weight of different contributions. 

Likewise one can use different kinematic variables to form \ot~odd moments. Instead of measuring the 
expectation values of the triple correlations among momenta listed above, one can measure the 
azimuthal angle between the $K \bar K$ and the $\pi^+\pi^-$ or $\mu^+\mu^-$ planes and search for a 
forward-backward asymmetry in it -- in analogy to what has been done for 
$K_L \to \pi^+\pi^-e^+e^-$ \cite{CPBOOK}. Without a specific model for the underlying \cp~odd dynamics one cannot decide a priori which correlations are most sensitive to such NP dynamics. 

\subsubsection{Benchmark Goals}

Viable NP scenarios could produce \cp~asymmetries close to the present experimental bounds, but 
not much higher. To have a realistic chance to find an effect, one should strive to reach 

$\bullet$ 
the ${\cal O}(10^{-4})$ [${\cal O}(10^{-3})$] level for time-dependent \cp~rate asymmetries in 
$D^0 \to K^+K^-$, $\pi^+\pi^-$, $K_S\rho^0$, $K_S\phi$ [$D^0 \to K^+\pi^-$]; 

$\bullet$ 
direct \cp~asymmetries in partial widths down to ${\cal O}(10^{-3})$ in $D\to K_S\pi$ and  in 
singly Cabibbo suppressed modes and down to ${\cal O}(10^{-2})$ in doubly Cabibbo suppressed 
modes; 

$\bullet$
the ${\cal O}(10^{-3})$ level in Dalitz asymmetries and \ot~odd moments.

\subsubsection{On Rare Charm Decays}

There seems to be general agreement that studying $D \to \gamma h/hh$ etc. is very unlikely to allow establishing the presence of NP because of uncertainties due to long distance dynamics 
\cite{BURD}. I am concerned that the same strong caveat applies also to $D \to l^+l^- h/hh$. 

The story is different for $D^0 \to \mu^+\mu^-$ which therefore deserves scrutiny. To get an independent 
estimate of SM long distance effects it would be very helpful, if one knew the width for 
$D^0 \to \gamma \gamma$. 

\section{Some Technical Explanations}
\label{TECH} 

The statements in the preceding Section are based on a whole body of technical considerations. It is 
not my intention to truly explain them here -- that would make this note rather unwieldy --, but just touch on a few relevant points to refresh the memory of the more experienced readers and to provide some guide to the literature for the rest. 

\subsection{On Theoretical Guidance}
\label{THGUID}

To say that theoretical predictions have not always been on the mark, in particular when nonperturbative forces were involved, is putting it very delicately. Yet this does not justify immediate rejection of theoretical advice -- it merely points to the need for some healthy skepticism.   Let me sketch two areas where theory can provide guidance already now or in the near future. 

Consider $D^0 \to K_S \pi^+\pi^-$, which has figured prominently in the signal for oscillations. Its Dalitz plot exhibits a rich structure with three classes of resonant final states in addition to its non-resonant 
contributions and their interferences: (i) Flavour-nonspecific transitions of the type 
$D^0 \to K_S \rho$, which are of particular interest for time dependent \cp~studies. 
(ii) Pure CA modes like $D^0 \to K^{*-}\pi^+$. (iii) Apparent DCS channels $D^0 \to K^{*+}\pi^-$, 
which can be populated also by $D^0 - \bar D^0$ oscillations. We do not have a theoretical description of the Dalitz plot that can be derived from first principles. Thus we have to rely on the next best thing -- models. One thing works in our favour here: once we have obtained a good fit to high statistics data with a model employing no more than a handful of free parameters, we can invoke the over-redundancy achieved with such a fit as a good measure of its quality. Based on the experience gained over the last twenty or so years we can state bounds that have to be obeyed by the ratios of DCS and CA decay amplitudes on rather general theoretical grounds. Such bounds then contribute to the aforementioned over-constraints. 

One can go one step further and argue that while a description of the full Dalitz plot based on 
first principles is well beyond our capabilities, we should be able to provide local descriptions, namely concerning the interference of a narrow resonance  with a broader distributions that to leading order can be assumed to be flat. To say it differently: we should be able to develop a framework for describing how the Dalitz plot 
population changes, when one starts on one side of a narrow resonance and then `moves' through it to the other side.

\subsection{On `Theoretical Engineering'}
\label{THEORENG}

When describing (quasi) two-body channels of the $D \to PP$, $PV$ type we can make an intelligent use of measured partial rates to get to a reliable theoretical description of \cp~asymmetries there. 
On each Cabibbo level one makes an ansatz for the total transition operator in terms of 
`elementary' $\Delta C = 1$ operators whose coefficients are computed from the known CKM factors and QCD radiative corrections. One makes a judicious choice of which such $\Delta C = 1$ operators to include -- corresponding to internal and external $W$ emission with or without interference, weak annihilation and Penguin contributions if possible. When evaluating the corresponding amplitudes one leaves the magnitude of the appropriate matrix elements and the values of their FSI phases open. From 
fitting such expressions to a comprehensive set of high statistics data one infers values for these a priori unknowns. The reliability of such extractions rests again on the degree of over-constraints one has achieved including cross referencing those numbers against each other using $SU(3)_{fl}$ relations etc. 

The ability to include also channels with (multi-)neutrals is obviously of essential value here, which belongs to the domain of $e^+e^-$ $\tau$-charm factories like BESIII \cite{BESBOOK}.  

\subsection{On \cpt~Constraints}
\label{CPT}

\cpt~symmetry provides more constraints than just equality of masses and lifetimes of particles and antiparticles \cite{CPBOOK}. For it tells us that the widths for {\em sub}classes of transitions have to be the same. For 
simplicity consider a toy model where the $D$ meson can decay only into two classes of final states 
$E=\{ e_i, i=1,...,n \}$ and $F=\{f_j, j=1,...,m \}$ with the strong interactions allowing members of the 
class $E$ to rescatter into each other and likewise for class $F$, but {\em no} rescattering possible 
{\em between} classes $E$ and $F$. Then \cpt~symmetry tells us partial width asymmetries 
{\em summed} over class $E$ already have to vanish and likewise for class $F$. This \cpt~`filter' can hardly be of any 
practical use for $B$ decays with their multitude of channels on vastly different CKM levels. 
Yet it might provide nontrivial validation checks for $D$ decays with their considerably fewer channels, where quasi-{\em elastic} unitarity could conceivably hold in a semiquantitative way. 

Penguins despite their poor reputation in flavour dynamics -- as expressed through the 
all too often heard "penguin pollution" -- are rather smart entities. When Penguin diagrams are invoked 
to generate the FSI required for a direct \cp~asymmetry, examining the light quarks in their loops will tell you in which class of channels the compensating asymmetries have to arise. To cite a simple example: such considerations suggest that a 
direct \cp~asymmetry in $D^0 \to K^+ K^-$ is compensated mainly by an asymmetry in 
$D^0 \to \pi^+ \pi^-$.  Finding these balancing effects would validate the observation of a presumably small asymmetry.

\subsection{On Relating Direct and Indirect Searches for New Physics}
\label{NPSCEN}

Looking for NP inducing \cp~violation is {\em not} a `wild goose chase'. For baryogenesis requires 
such dynamics. Together with the other advantages offered by $D$ decays this provides strong 
motivation for dedicated searches. Those can be seen as `hypothesis-generating' research -- 
similar to the {\em present} situation in $B$ physics. Once, say, SUSY is found in high 
$p_{\perp}$  collisions at the LHC, future studies of $B$ decays would be of the 
`hypothesis-probing' variety like about fifteen years ago, when the $e^+e^-$ $B$ factories were approved. Finding direct evidence for LHT models might turn the same trick for the detailed study of 
$D$ decays.

\section{Summary and Outlook}
\label{OUT}

There is general conviction that $D^0 - \bar D^0$ oscillations have been observed:  
$(x_D,y_D) \neq (0,0)$. However their theoretical interpretation is rather ambiguous: 
SM dynamics might generate the whole effect or the major part of it or only a minor part. 
Deciding this issue on theoretical grounds would require a breakthrough in our computational 
abilities. A comprehensive and detailed program of \cp~studies in charm transitions can presumably 
decide the issue: It might establish the intervention of New Physics; for even if New Dynamics provided 
only a non-leading contribution to $\Delta M_D$, they would quite possibly represent the leading source of 
\cp~asymmetries due to the `dullness' of SM \cp~phenomenology. The present absence of any signal is 
not very telling. For future studies we need to know the relative size of 
$x_D$ and $y_D$ as best as possible. 

`Realistically' one cannot hope for much more than ${\cal O}(10^{-3})$ effects. Thus we have to learn to 
exploit the statistical `muscle' of LHCb and control systematics. Asymmetries in final state distributions as analyzed through Dalitz studies and \ot~odd correlations offer several advantages: differential asymmetries could be considerably larger than integrated ones; internal cross checks provide powerful tools to deal with systematics; they can provide us with novel clues about the nature of the intervening New Physics. 
On the 
theory side we can expect a positive learning curve for theorists, yet should not count on 
miracles.

\vspace{0.5cm}

{\bf Acknowledgments:} I have benefitted from several discussions with colleagues from the LHCb 
collaboration. This work was supported by the NSF under the grant number PHY-0807959.

\vspace{4mm}


\end{document}